\documentclass[11pt]{article}
\usepackage{moriond,epsfig,subfigure}
\usepackage{cite,mcite}
\usepackage{hyperref}
\usepackage{multirow}
\usepackage{amsfonts,amsmath,amssymb}

\def\Journal#1#2#3#4{{#1} {\bf #2}, #3 (#4)}

\def\PRD{{\em Phys. Rev.} D}

\def\be{\begin{equation}}
\def\ee{\end{equation}}
\def\bea{\begin{eqnarray}}
\def\eea{\end{eqnarray}}

\renewcommand{\d}{\mathrm{d}}
\def\dndeta{\ensuremath{\d N_{\textrm{ch}}/\d \eta}}

\def\lumi#1{\ensuremath{10^{#1}~\textrm{cm}^{-2}\textrm{s}^{-1}}}
\def\rts {\ensuremath{\sqrt{s}}}
\def\pt{\ensuremath{p_T}}

\def\et{\ensuremath{E_T}}

\def\TeV{\ifmmode {\mathrm{\ Te\kern -0.1em V}}\else
                   \textrm{Te\kern -0.1em V}\fi}%
\def\GeV{\ifmmode {\mathrm{\ Ge\kern -0.1em V}}\else
                   \textrm{Ge\kern -0.1em V}\fi}%
\def\MeV{\ifmmode {\mathrm{\ Me\kern -0.1em V}}\else
                   \textrm{Me\kern -0.1em V}\fi}%

\def\ipb{\mbox{pb$^{-1}$}}%  Inverse picobarns.

\def\etal{{\rm et~al.}}%

\begin{document}
\vspace*{3cm}
\title{MINIMUM BIAS AND UNDERLYING EVENT STUDIES\\AT ATLAS AND CMS}

\author{M. LEYTON\\(on behalf of the ATLAS and CMS Collaborations)}

\address{Lawrence Berkeley National Laboratory, 1 Cyclotron Rd,\\
Berkeley CA 94720, USA}

\maketitle\abstracts{
An overview of minimum bias and underlying event studies at the LHC with the ATLAS and CMS detectors is presented. Current uncertainties in the modeling of soft $pp$ inelastic interactions at the LHC energy scale are discussed. Triggers used to select inelastic interactions at ATLAS and CMS are described and compared. A summary of some of the ongoing minimum bias and underlying event analyses by the ATLAS and CMS collaborations is given.}

Minimum bias and underlying event properties have previously been studied over a wide range of energies. In particular, results from experiments at CERN and Fermilab have been used to tune the \textsc{Pythia}~\cite{pythia} and \textsc{Phojet}~\cite{phojet} Monte Carlo event generators. However, due to uncertainties in the modeling of the energy dependence of soft inelastic interactions, these generators give widely varying predictions when extrapolated to LHC energies~\cite{moraes}. 

\section{LHC, ATLAS \& CMS}

The Large Hadron Collider (LHC)~\cite{lhcPaper} at CERN will collide protons at a center-of-mass energy of $\rts = 14 \TeV$. An initial physics run at $10 \TeV$ is expected to start in October 2009. Peak luminosity for the 2009--2010 run will be $\mathcal{L} = 5 \times \lumi{31}$, or about 1.8 interactions per bunch crossing. Total integrated luminosity during the first 100 days will be about $100~\ipb$.

ATLAS~\cite{atlasPaper} and CMS~\cite{cmsPaper} are $4\pi$ general-purpose detectors designed for high-luminosity studies at the LHC. Tracking detectors for both ATLAS and CMS cover a pseudorapidity range of $|\eta| < 2.5$. Both experiments have a multi-layer silicon tracker with pixels and strips. In addition, ATLAS has a transition radiation straw tracker. The ATLAS and CMS tracking chambers are immersed in a solenoidal magnetic field of 2 T and 4 T, respectively. Both detectors have electromagnetic calorimetry covering $|\eta| \lesssim 3.0$ and hadronic calorimetry covering $|\eta| \lesssim 5$.

\section{Minimum Bias}

`Minimum bias' events are usually associated with the non-single-diffractive (NSD) inelastic portion of the total cross section.\footnote{The total cross section ($\sigma_{\textrm{tot}}$) at the LHC can be written as a sum of elastic ($\sigma_{\textrm{elas}}$) and inelastic ($\sigma_{\textrm{inel}}$) components, with the inelastic part further sub-divided into single-diffractive ($\sigma_{\textrm{sd}}$), double-diffractive ($\sigma_{\textrm{dd}}$) and non-diffractive ($\sigma_{\textrm{nd}}$) components~\cite{moraes}. Central diffraction has been ignored.} They are dominated by soft interactions, with low transverse momentum (\pt) and low particle multiplicity. Studies with minimum bias events are possible with very early data (1--$10~\ipb$), ideally during low-luminosity running, when the number of $pp$ collisions per bunch crossing is $\leq 1$. At higher luminosities, minimum bias will be a major background, with an average of 18 minimum bias interactions per bunch crossing at LHC design luminosity~\cite{cscBook}. It is therefore very important to have an accurate model of minimum bias for all other high-\pt\ physics measurements.

A list of some of the important observables in minimum bias events, along with corresponding predictions by \textsc{Pythia} version 6.214 (ATLAS tune) and \textsc{Phojet} version 1.12, is given here~\cite{moraes}:
\begin{itemize}
\setlength{\itemsep}{6pt}
\item total cross section, $\sigma_{\textrm{tot}}$: 102--119 mb (17\% difference); 
\item average charged particle multiplicity, $\langle n_{\textrm{ch}} \rangle$: 119--70 (30\% difference);
\item density of charged particles in the central region, $\displaystyle \dndeta |_{\eta=0}$: 6.8--5.1 (30\% difference);
\item average transverse momentum, $\langle \pt \rangle$: 0.55--$0.64 \GeV/c$ (17\% difference).
\end{itemize}
Measuring \dndeta\ (see section~\ref{sec:dndeta}) at different LHC collision energies will be especially crucial for choosing between models since \textsc{Pythia} and \textsc{Phojet} predict drastically different behaviors with different logarithmic dependencies~\cite{moraes}. The average \pt\ in the central region is also important since it is very sensitive to the tuning of Multiple Partonic Interactions (MPI)~\cite{field}. 

Because of the wide range in the predictions of these quantities, a first measurement to within 10\% can discriminate between the different models.

\subsection{Minimum bias trigger}

A minimum bias trigger is intended to select inelastic collisions with as little bias as possible. Several types of minimum bias triggers have been implemented by ATLAS and CMS in order to cover various phases of LHC running~\cite{cscBook, pas_trigger}.

A \textit{random trigger} (with beam pickup) at Level-1 is ideal since it accepts all types of inelastic events equally with zero bias. However, random triggering is very inefficient at low luminosities ($\mathcal{L} < \lumi{30}$) since the probability of an interaction during a bunch crossing is $< 1\%$. At higher trigger levels, signals in the tracking detectors can be used to suppress empty bunch crossings (noise events). This so-called \textit{track trigger} selects events with a minimum number of silicon hits or reconstructed tracks or both. These two trigger strategies are often combined into a random-based track trigger~\cite{cscBook}.

CMS will also use their forward calorimeter (FCAL), located between $3.0 < |\eta| < 5.0$, as a minimum bias trigger by requiring a minimum number of towers with transverse energy (\et) greater than some threshold. ATLAS will use the Minimum Bias Trigger Scintillators (MBTS), located between $2.1 < |\eta| < 3.8$, by requiring energy deposit above threshold in one or more counters. Various scenarios and thresholds have been studied by both collaborations~\cite{cscBook, pas_trigger}.

A summary of the ATLAS and CMS minimum bias triggers discussed here is given in table~\ref{table:mbtriggers}. Trigger efficiencies are shown for the various physics processes. All but one of the triggers are highly efficient at selecting non-diffractive events. The diffractive events, however, have much lower trigger efficiencies since these events typically have a lower track multiplicity than non-diffractive events, especially in the central tracking region. It is important to note that neither ATLAS nor CMS have a pure NSD trigger. Model-dependent corrections are therefore needed to compare measured distributions to past experiments~\cite{cscBook}.
\begin{table}[htbp]
  \caption{Efficiencies of selected ATLAS and CMS minimum bias triggers for various physics processes simulated at $\rts = 14 \TeV$. An \et-threshold of $1 \GeV$ has been used in the CMS FCAL trigger study.
    \label{table:mbtriggers}}
  \vspace{0.4cm}
  \begin{center}
    \begin{tabular}{|c|c|c|c|c|c|}
      \hline
      {\small Experiment} & {\small Trigger} & {\small Non-diff.} & {\small Double-diff.} & {\small Single-diff.} & {\small Beam gas} \\ \hline
      \multirow{3}{*}{\small ATLAS} & {\small Track ($\geq 2$ tracks)} & {\small 100\%} & {\small 65\%} & {\small 57\%} & {\small 40\%} \\
      & {\small MBTS (2 hits anywhere)} & {\small 100\%} & {\small 83\%} & {\small 69\%} & {\small 54\%} \\
      & {\small MBTS (1 hit on each side)} & {\small 99\%} & {\small 54\%} & {\small 45\%} & {\small 40\%} \\ \hline
      \multirow{3}{*}{\small CMS} & {\small Track ($\geq 1$ track)} & {\small 99\%} &{\small  69\%} & {\small 59\%} & -- \\
      & {\small FCAL (1 tower on one side)} & {\small 81\%} & {\small 15\%} & {\small 15\%} & -- \\
      & {\small FCAL (1 tower on each side)} & {\small 48\%} & {\small 1\%} & {\small 1\%} & -- \\ \hline
    \end{tabular}
  \end{center}
\end{table}

\subsection{Measuring the pseudorapidity density}
\label{sec:dndeta}

One of the first physics results from ATLAS and CMS will be the measurement of the pseudorapidity density of charged particles, \dndeta, in minimum bias events at the LHC. This analysis has been prepared by both experiments using a full detector simulation~\cite{cscBook,pas_dndeta}. The ATLAS version uses the MBTS trigger to define a minimum bias event sample. A selected subset of fully reconstructed tracks are then counted and corrected back to MC generator level. Track and vertex reconstruction inefficiencies are taken into account. The CMS version uses a random trigger, thereby selecting inelastic events. Hits are counted in each layer of the pixel detector and a conversion function is used to map these hits into tracks in pseudorapidity. Corrections are applied for event selection and looping low-\pt\ tracks. Neither the ATLAS nor the CMS analysis corrects for bias imposed by the trigger.

\begin{figure}[htbp]
  \begin{center}
    \includegraphics[height=6.9cm]{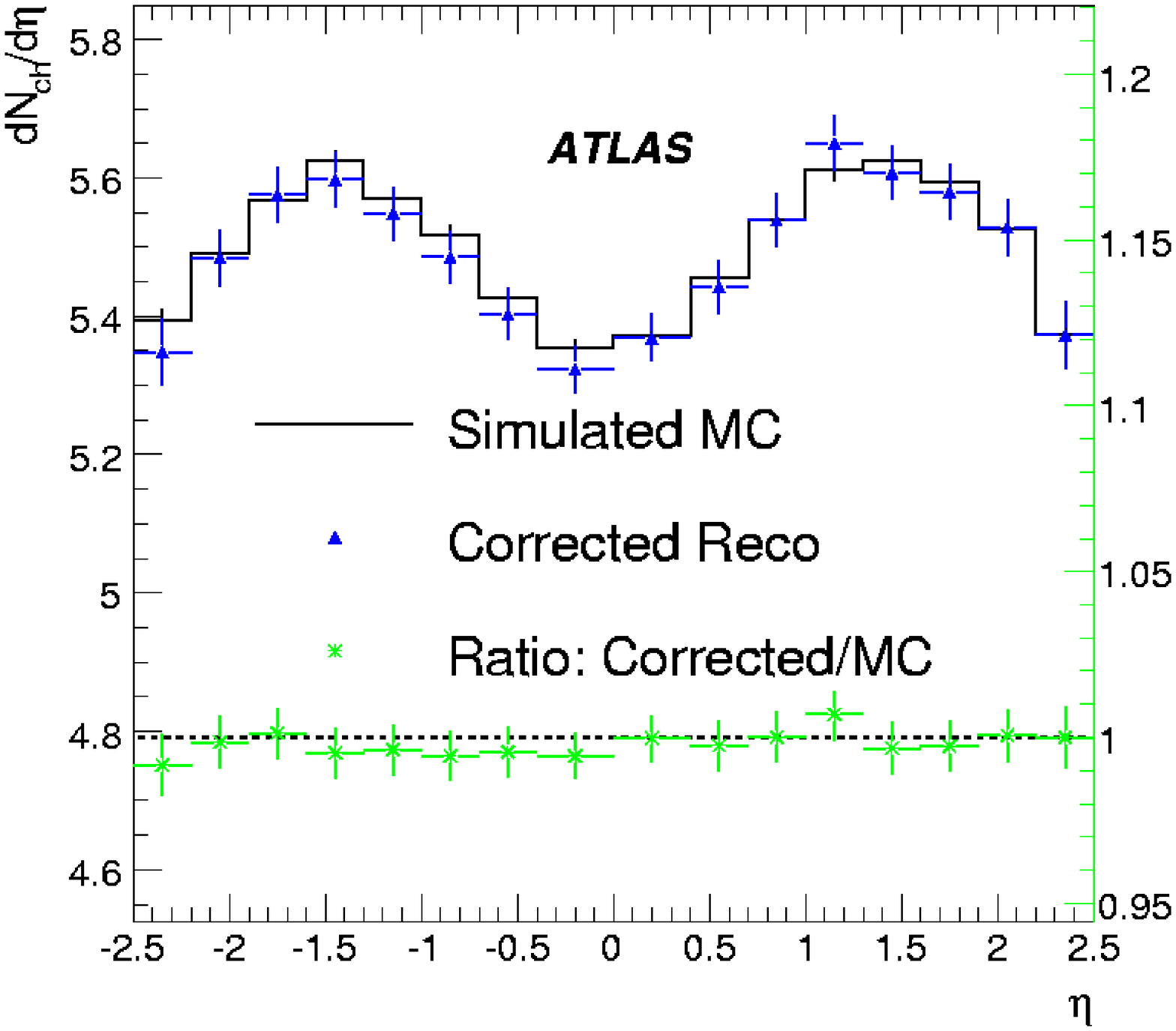}
    \includegraphics[height=6.9cm]{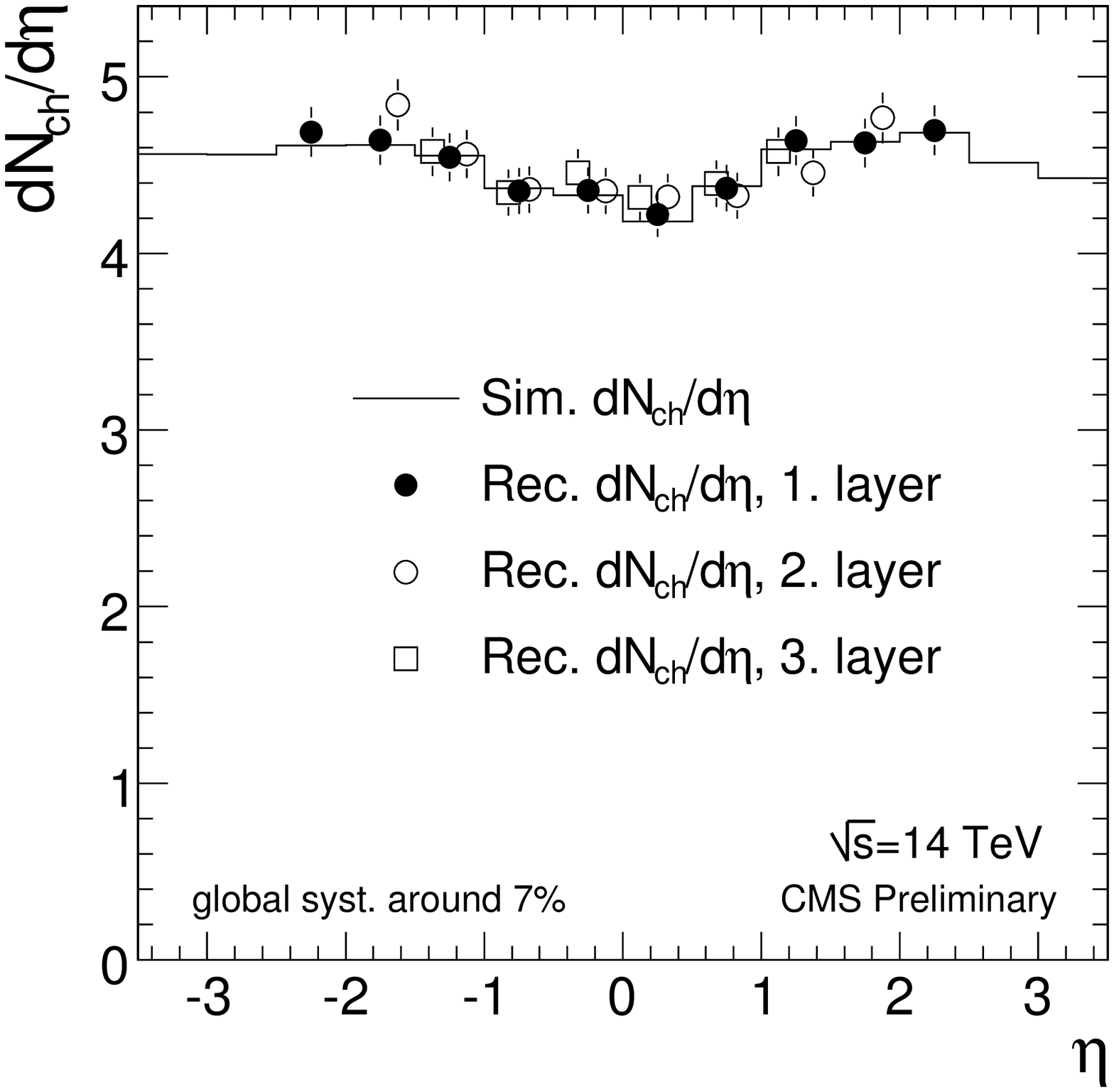}
  \end{center}
  \caption{Results of simulated \dndeta\ analyses for the ATLAS minimum bias (\textit{left}) and the CMS inelastic (\textit{right}) event samples. Distributions have been integrated over $\pt > 150 \MeV/c$ (\textit{left}) and $\pt > 30 \MeV/c$ (\textit{right}).} 
\label{figure:dndeta}
\end{figure}

Figure~\ref{figure:dndeta} shows the results of the two analyses, using events generated with \textsc{Pythia} at $\rts = 14 \TeV$. The generated \dndeta\ distributions for the minimum bias (ATLAS) and inelastic (CMS) event sample have been reconstructed. The inelastic sample used by CMS includes a greater fraction of single and double-diffractive events which lower the average track multiplicity in the central pseudorapidity region. A lower \pt\ range is accessible by CMS ($\pt > 30 \MeV/c$, compared to $\pt>150 \MeV/c$ for ATLAS) since no tracking or alignment is needed by the hit-counting method.

The total estimated systematic uncertainty for both analyses is about 7\%. The dominant source of uncertainty in the ATLAS analysis comes from mis-alignment of the tracker, but this is expected to decrease once the alignment from cosmic ray data is considered. The dominant sources of systematic uncertainty at CMS come from vertex bias and the hit-to-track conversion function.

\section{Underlying Event}

The Underlying Event (UE) is everything in the event except the hard scatter, including Initial State Radiation (ISR), Final State Radiation (FSR) and hadronization and beam remnants. From an experimental point of view, it is impossible to separate these two things; however, topological properties of the event can be used to define a set of physics observables which are sensitive to different aspects of the UE~\cite{pythia_tunes}. Studying the UE is important for understanding the evolution of QCD with collision energy as well as understanding the systematic corrections on many studies, from basic detector calibrations to physics analyses such as mass measurements. 

Jet events are ideal for studying the UE. The approach by CDF is to look at observables in $\eta$-$\phi$ regions transverse (between 60$^\circ$ and 120$^\circ$ in azimuth) to the high-\pt\ objects (\textit{leading jets})~\cite{affolder}. Important observables include particle density, energy density and $\sum \pt$. The CDF analyses have shown that the density of particles in the UE in jet events is about a factor of two larger than in a typical minimum bias collision, a feature known as the `pedestal effect'~\cite{pythia_tunes}. This tells us that the UE is not the same as minimum bias but that they are closely related.

As in the case of minimum bias, MC predictions at LHC energies are very uncertain, at the level of several hundred percent. For example, the particle density of events with leading jet $\pt > 10 \GeV/c$ is predicted to be between 13 and 29, a difference of over 200\%. Uncertainties for underlying event predictions at the LHC come from the parton density functions, ISR/FSR gluon radiation, color flow and the modeling of MPI. Because of the wide range in LHC predictions, we can expect that 2009--2010 LHC data can already discriminate between models. Constraining of models and tuning of generators can follow shortly thereafter.

\vskip 0.5cm

\begin{figure}[htbp]
  \begin{center}
    \includegraphics[height=6.9cm]{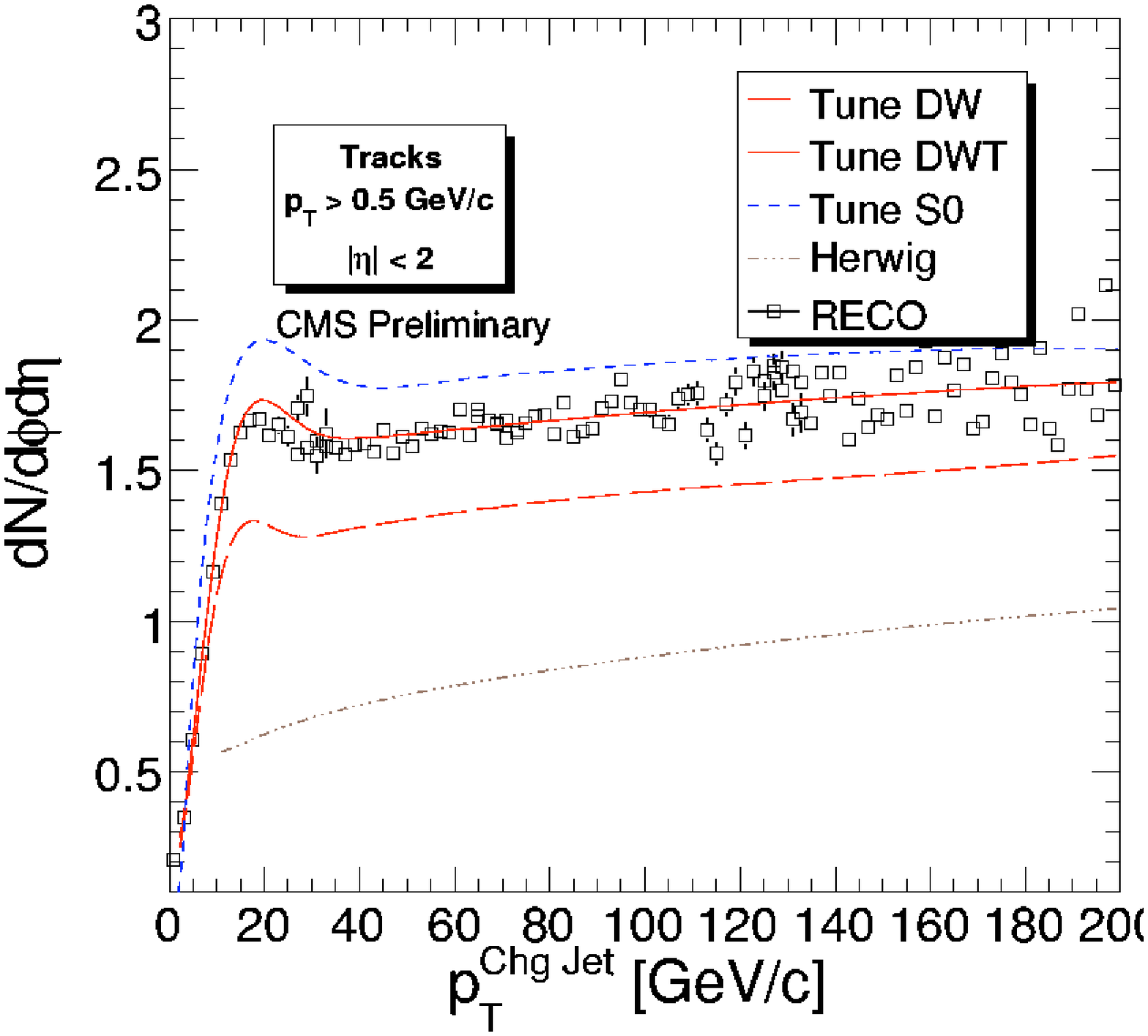}
    \includegraphics[height=6.9cm]{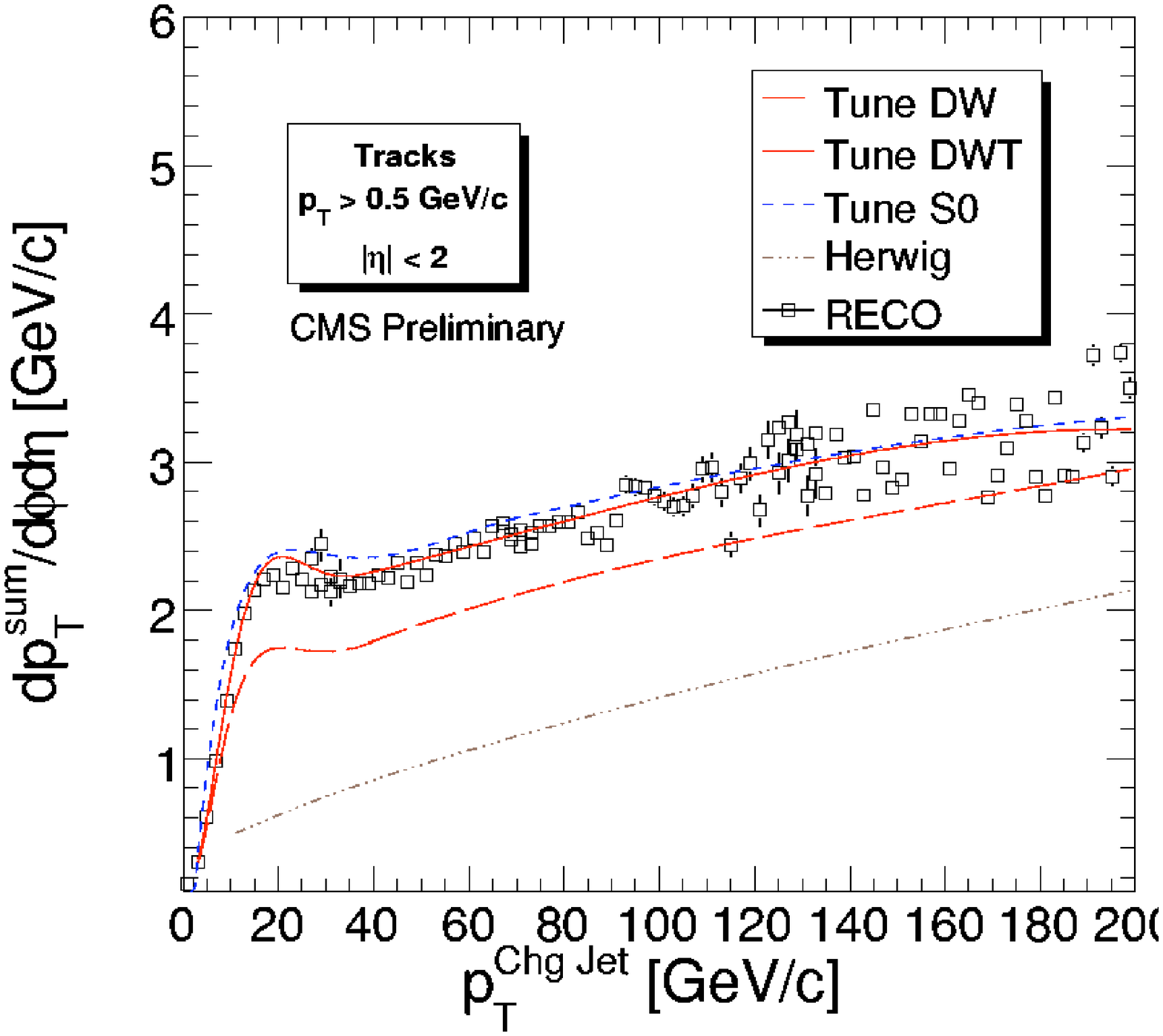}
  \end{center}
  \caption{Results of a simulated CMS underlying event study. Corrected densities \dndeta$\d \phi$\ (\textit{left}) and $\d \pt^{\textrm{sum}}/\d \eta \d \phi$\ (\textit{right}) for tracks with $\pt > 0.5 \GeV/c$, as a function of the leading charged jet $\pt$, in the transverse region, for an integrated luminosity of $100~\ipb$.}
\label{figure:cms_ue}
\end{figure}

\noindent \textbf{Measuring the underlying event:} Figure~\ref{figure:cms_ue} shows the results of a simulated underlying event study by CMS using 100 \ipb\ of data~\cite{pas_ue}. Events were selected using a minimum bias trigger and an additional trigger on the \pt\ of the leading jet (20, 60 and $120 \GeV/c$). The plots show the MC generated curves for track multiplicity and $\sum \pt$\ per $\d \eta \d \phi$\ region as a function of the charged jet \pt. The curve corresponding to the \textsc{Pythia} DWT tune has been reconstructed by the analysis for $\pt > 0.5 \GeV/c$. With this much data, the reconstruction can resolve the various \textsc{Pythia} tunes quite well. A summary of the parameters for each of the various \textsc{Pythia} tunes considered in this study can be found in~\cite{pythia_tunes}.

\section{Conclusion}

The revised LHC program will include an initial physics run at $\rts = 10 \TeV$, followed by running at $14 \TeV$. This will allow measuring the energy dependence of Minimum Bias and Underlying Event, which is important for model tuning and constraining. Although current models of minimum bias and the underlying event diverge on LHC predictions, $1 < \int \mathcal{L} \d t < 10~\ipb$ of data is sufficient to discriminate between these models. ATLAS and CMS are both ready to study minimum bias and the underlying event at the LHC. Reconstructed distributions can be accurately corrected back to MC level. These analyses are an important first step in finding a model that describes the entire energy range up to the LHC.

\section*{Acknowledgments}

I would like to thank the ATLAS and CMS collaborations for their contributions and the speakers committees for this opportunity. I also thank the Moriond program and organizing committees for perpetuating the \textit{spirit of Moriond}. This work was supported by the U.S. National Science Foundation.

\end{document}